\documentclass[reprint,
superscriptaddress,
 amsmath,amssymb,
 aps,
]{revtex4-2}

\usepackage{graphicx}
\usepackage{dcolumn}
\usepackage{bm}
\usepackage[colorlinks=true, linkcolor=blue, urlcolor=blue,citecolor=blue]{hyperref}

\usepackage{xcolor}
\begin{document}

\preprint{APS/123-QED}




\title{Giant spin Nernst effect in a two-dimensional antiferromagnet due to  magnetoelastic coupling-induced gaps and interband transitions between magnon-like bands}

\author{D.-Q. To}%
\affiliation{Department of Materials Science and Engineering, University of Delaware, Newark, DE 19716, USA}%

\author{C. Y. Ameyaw}
\affiliation{Department of Materials Science and Engineering, University of Delaware, Newark, DE 19716, USA}%

\author{A. Suresh}
\affiliation{Department of Physics and Astronomy, University of Delaware, Newark, DE 19716, USA}%

\author{S. Bhatt}
\affiliation{Department of Physics and Astronomy, University of Delaware, Newark, DE 19716, USA}%

\author{M. J. H. Ku}
\affiliation{Department of Materials Science and Engineering, University of Delaware, Newark, DE 19716, USA}%
\affiliation{Department of Physics and Astronomy, University of Delaware, Newark, DE 19716, USA}%

\author{M. B. Jungfleisch}%
\affiliation{Department of Physics and Astronomy, University of Delaware, Newark, DE 19716, USA}%

\author{J. Q. Xiao}%
\affiliation{Department of Physics and Astronomy, University of Delaware, Newark, DE 19716, USA}%

\author{J. M. O. Zide}%
\affiliation{Department of Materials Science and Engineering, University of Delaware, Newark, DE 19716, USA}%

\author{B. K. Nikoli\'{c}}%
\email{bnikolic@udel.edu}
\affiliation{Department of Physics and Astronomy, University of Delaware, Newark, DE 19716, USA}%

\author{M. F. Doty}%
 \email{doty@udel.edu}
\affiliation{Department of Materials Science and Engineering, University of Delaware, Newark, DE 19716, USA}%

\date{\today}

\begin{abstract}
We analyze theoretically the origin of the  spin Nernst and thermal Hall effects in FePS$_3$ as a realization of two-dimensional antiferromagnet (2D AFM). We find that a strong magnetoelastic coupling, hybridizing magnetic excitation (magnon) and elastic excitation (phonon), combined with time-reversal-symmetry-breaking, results in a Berry curvature hotspots in the region of anticrossing between the two distinct hybridized bands. Furthermore,  large  spin Berry curvature emerges due to {\em interband transitions between two magnon-like bands}, where a small energy gap is induced  by magnetoelastic coupling between such bands that are {\em energetically distant} from anticrossing of hybridized bands. These nonzero Berry curvatures generate topological transverse transport (i.e., the thermal Hall effect) of hybrid excitations, dubbed magnon-polaron, as well as of spin  (i.e., the spin Nernst effect) carried by them, in response to applied longitudinal temperature gradient. We investigate the dependence of the spin Nernst and thermal Hall conductivities on the applied magnetic field and temperature, unveiling very large spin Nernst conductivity {\em even} at zero magnetic field. Our  results suggest FePS$_3$ AFM, which is already available in 2D form experimentally,  as a promising platform to explore the topological transport of the magnon-polaron quasiparticles at THz frequencies.   
\end{abstract}

\maketitle

\section{Introduction}

Two-dimensional (2D) antiferromagnets (AFMs~\cite{Gibertini2019} are attracting growing attention due to their potential application as material platforms for spintronics, spin-orbitronics, and spin-caloritronics \cite{Jungfleisch2018,Hao2018,Huang2020,Dolui2020,Yang2021,Jiang2021,Tan2021,Phan2021,Xiong2022}. Because the strong exchange interaction between their localized spins results in intrinsic THz frequency dynamics, AFMs are particularly promising for the development of devices with high operating speeds. For example, magnon in a 2D AFM can be employed to store and transfer THz frequency information without Joule heating due to the absence of a charge current or a stray field. Such materials can also provide efficient spin-transport channels in spintronic devices with low energy consumption \cite{Suresh2020,Zhang2020a,Suresh2021, Bajpai2021,Belvin2021,Chumak2022}. Despite these advantages, the use of magnons in 2D AFMs as a part of realistic devices is severely limited by the lack of efficient ways to generate and manipulate magnon excitations. The hybridization of magnons and phonons may provide a path toward coherent control of magnons in 2D AFM material via a manipulation of the hybridized states \cite{Li2020a,Yahiro2020,Vidal2020,Tabatabaei2021,Kikkawa2022}. For instance, it has been shown that one can electrically generate magnon spin current through the interaction between magnon and phonon \cite{Chen2015,Nomura2019}. Conversely, it has also been shown that the dynamics of a phonon can be controlled via its interaction with a magnon \cite{Zhao2020,Holanda2021,Rezende2021}.

\begin{figure}[h!]
\centering
    \includegraphics[width=.5\textwidth]{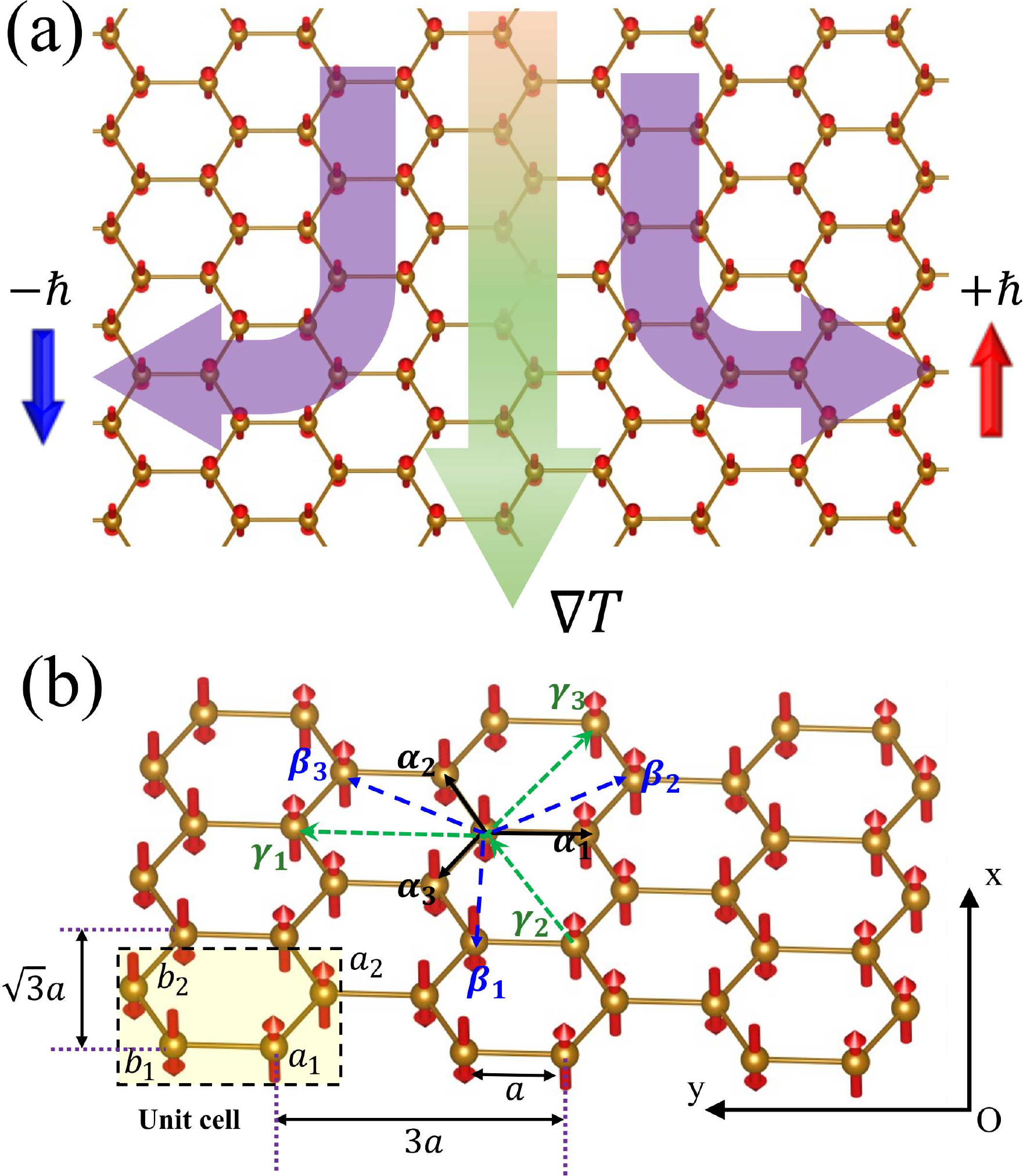}
 \caption{(a) Schematic view of the magnon SNE in a 2D AFM where  transverse flow of magnons carrying opposite out-of-the plane spins ($\pm ~\hbar$) is induced by temperature gradient $\nabla T$ along the longitudinal direction~\cite{Zhang2022}. (b) The quasi-2D lattice of FePS$_3$ formed by Fe atoms. The arrows indicate the direction of the its localized spins within zigzag AFM phase considered in our study. Here $\alpha_{i}$, $\beta_{i}$ and $\gamma_{i}$ $(i=1,2,3)$ are the vectors joining the first, second, and third-nearest neighbors, respectively. A unit cell contains four Fe atoms forming a rectangularly-shaped BZ with periodicity in real space that is $\sqrt{3}a$ or $3a$ long in the $x$- or $y$-directions (where $a$ is the lattice constant), respectively.}
  \label{Structure}
\end{figure}

Magnons and phonons are the collective and charge-neutral excitations of localized spins and lattice vibrations, respectively. They behave as bosonic quasiparticles, obeying the Bose-Einstein distribution function at finite temperature with zero chemical potential in equilibrium due to their non-conserved number. Strong coupling between a magnon and a phonon results in a hybridized state that includes both spin and lattice collective excitations in a single coherent mode~\cite{Park2020,Zhang2020,Bazazzadeh2021,Go2022}. As a result, a new type of quasiparticle, dubbed magnon-polaron~\cite{Zhang2021,Luo2023}, is formed. The intriguing and non-trivial emergent properties of magnon-polarons provide a possible foundation for novel devices with unique optical and electrical functionalities \cite{Xu2018,Godejohann2020,Berk2020,Bozhko2020,Awschalom2021,Li2021,Mertens2022}. In particular, the hybridization of magnons and phonons to create a magnon-polaron can  generate a finite Berry and spin (generalized)  Berry curvatures  concentrated around anticrossing regions~\cite{Park2020,Zhang2020,Bazazzadeh2021,Go2022} of the magnon and phonon bands. These Berry curvatures then lead to nontrivial topological transverse transport---the magnon thermal Hall effect (THE) and magnon spin Nernst effect (SNE)---which have attracted a lot of attention~\cite{Zyuzin2016,Takahashi2016,Murakami2017,Go2019,Zhang2019,Okamoto2020,Park2020,Zhang2020,Bazazzadeh2021,Li2020,Go2022,Ma2022,Zhang2022,Luo2023}. In particular, recent studied have demonstrated~\cite{Zhang2021,Liu2021,Vaclavkova2021,Pawbake2022,Sun2022} possibly strong magnon-phonon coupling  FePS$_3$ as the realization of 2D AFM. This, together with experimentally accessible 2D form of this material, makes FePS$_3$ a great candidate for investigation of magnon THE and SHE. 


Let us recall that the magnon THE~\cite{Murakami2017}  refers to a phenomenon that occurs when a temperature gradient applied to a magnetic material generates transverse thermal transport of magnons, perpendicular to both the temperature gradient and magnetization. The magnon SNE~\cite{Zhang2022}, as an analogy of the electronic spin Hall effect (SHE) ~\cite{Sinova2015,Nikolic2006} where electrons of opposite spin travel in opposite directions transverse to applied unpolarized charge current, involves the flow of magnons instead of electrons carrying opposite spin flow in opposite directions perpendicular to the temperature gradient [Fig.~\ref{Structure}(a)]. The magnon SNE is made possible by the existence of two magnon species within AFM carrying opposite spin polarization~\cite{Zhang2022}. Recent studies have shown that the magnon SNE effect can be observed in: collinear antiferromagnets~\cite{Zhang2022,Cheng2016,Zyuzin2016} on a honeycomb lattice, where the Dzyaloshinskii-Moriya interaction (DMI) acting~\cite{Hog2017} on magnons plays an analogous role as  spin-orbit coupling (SOC) plays~\cite{Sinova2015,Nikolic2006} for electrons in the SHE; noncollinear antiferromagnets~\cite{Mook2019,Li2020}, even without any SOC  responsible for DMI, and in zero applied magnetic field; as well as in collinear antiferromagnets~\cite{Zhang2020, Bazazzadeh2021,Go2022} or ferrimagnets~\cite{Park2020} with magnetoelastic coupling hybridizing magnon and phonon quasiparticle bands whose anticrossing regions are putatively crucial~\cite{Park2020} to obtain nonzero Berry and spin Berry curvature driving [see Sec.~\ref{sec:conductivities}] transverse transport in THE and SNE, respectively.

In contrast, our study highlights a mechanism~\cite{Go2022} where a significant spin Berry curvature can be induced in an energy window of magnon-like bands that is {\em energetically distant} [for example the 1st and 2nd band in Fig.~\ref{Bands}(a)] from the magnon-phonon hybridized bands and their anticrossing within a collinear AFM. The magnon-like bands posses a small phonon character [Fig.~\ref{Bands}(a)] over the entire Brillouin zone (BZ), which causes opening of slight band gaps between them [Fig.~S2(b) in SM~\footnotemark[1]]. These band gaps are actually {\em smaller} than anticrossing gap between magnon-like and phonon-like bands [Fig.~S2(b) in SM~\footnotemark[1]]. The {\em smallness} of band gaps between magnon-like bands [Figs.~S2(b) and S3(b) in SM~\footnotemark[1]] and phonon-mediated interband transitions~\cite{Go2022} between them lead to significant spin Berry curvature (Fig.~\ref{Spinberry}) and, thereby, the possibility of a giant SNE in FePS$_3$ collinear AFM. 

The paper is organized as follows. In Sec. \ref{theoryandmodel} we introduce an effective Hamiltonian to capture the magnon-phonon hybridization  within 2D AFMs belonging to the MPX$_3$ (M = Fe, Mn, Co, Ni; X = S, Se) family hosting localized spins and their magnetic moments in a zigzag phase. The same Section also  reviews the theoretical framework of linear-response theory that can be used to investigate the transverse transport of magnon-polaron quasiparticles. In Sec.~\ref{results} we discuss thus generated SNE and THE for FePS$_3$, including  the dependence of the thermal Hall and spin Nernst conductivities on the applied magnetic field and temperature. We conclude in Sec.~\ref{conclu}.

\section{Models and Methods}
\label{theoryandmodel}

\subsection{2D AFM Hamiltonian describing magnons, phonons and their magnetoelastic coupling}

The MPX$_3$ (X = Fe, Mn, Co, Ni; X = S, Se) family of materials are van der Waals magnets~\cite{Gibertini2019} forming layered structures that are weakly bound by van der Waals forces and possess a stable magnetic order even in the monolayer limit~\cite{Olsen2019,Vanherck2020} because of a huge single ion anisotropy energy \cite{Olsen2021,Du2016,Lee2016,Liu2021,Luo2023,Cui2023}. In particular, Fig.~\ref{Structure} shows the layered  structure of FePS$_3$ that is established solely by the Fe atoms. Within each layer, the Fe atoms arrange in a honeycomb-like lattice structure with ``columns" of spins having opposite spin moments. We consider the FePS$_3$ magnetic structure in the so-called zigzag AFM phase in which a unit cell contains two pairs of equivalent atoms (i.e., having the same spin direction) that are labelled as $a_{i}$ and $b_{i}$ $(i=1,2)$, respectively. Due to the small value of the interlayer exchange interaction relative to the intralayer exchange interaction, these AFMs are, to a very good approximation, quasi-two dimensional magnets even in the bulk \cite{Wildes2012,Lancon2016,Olsen2021,To2022,Basnet2022,Zollner2022}. The magnon-phonon hybridization in FePS$_3$ can, therefore, be  investigating by focusing on quasi-2D honeycomb structure of Fe atoms whose Hamiltonian can be written as
\begin{equation}\label{Ham_total}
    H = H_{m} + H_{p} + H_{mp}.
\end{equation}
Here $H_{m}$ is the Hamiltonian of localized spins whose low-energy excited states are magnons~\cite{Bajpai2021}; $H_{p}$ is the phonon Hamiltonian; and $H_{mp}$ is the term describing magnetoelastic coupling and thereby induced hybridization of magnons and phonons. 
\begin{figure}
\centering
    \includegraphics[width=0.495\textwidth]{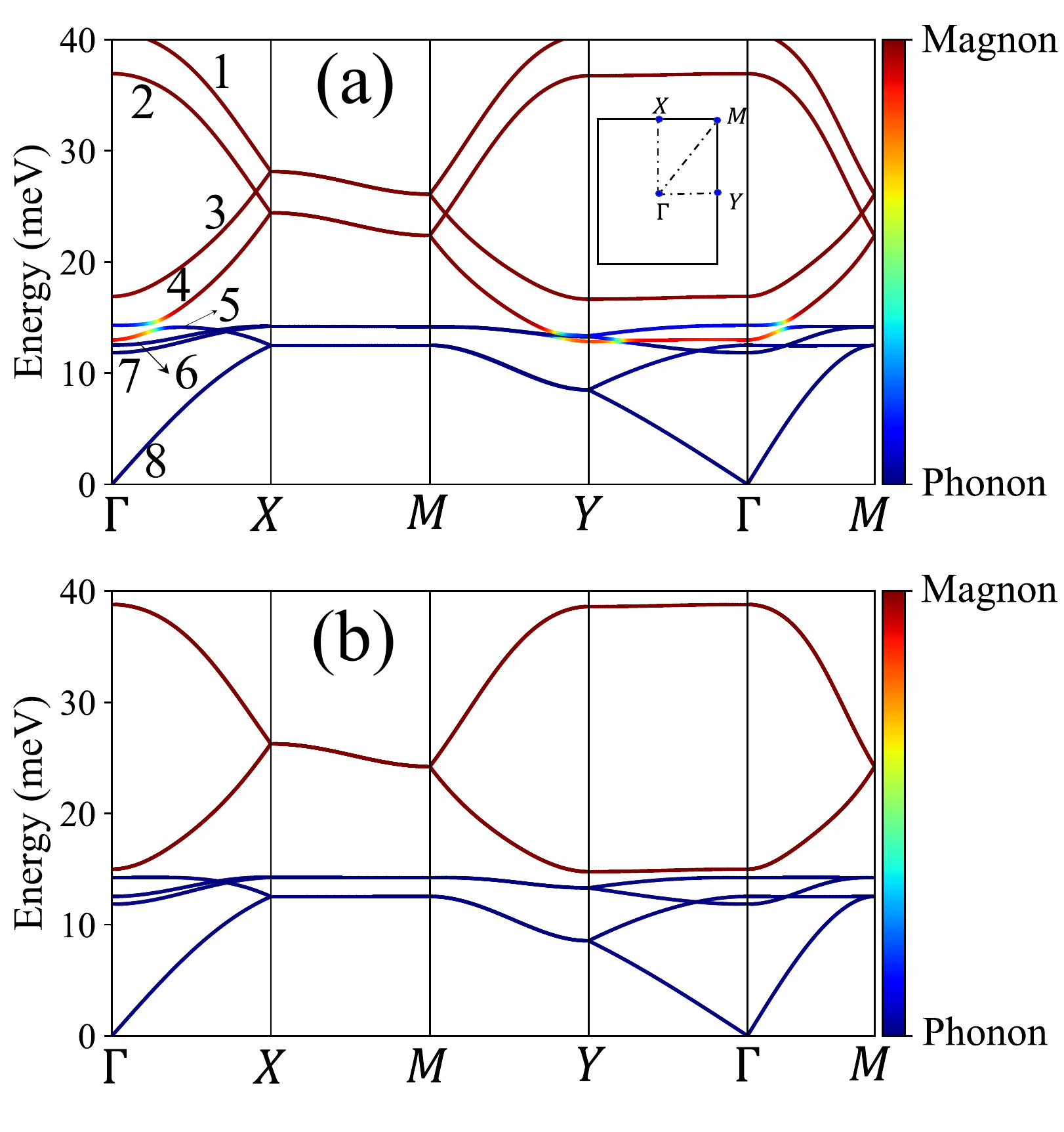}
 \caption{(a) The hybridized magnon-phonon band structure of FePS$_3$ [Fig.~\ref{Structure}], along \mbox{$\Gamma$-$X$-$M$-$Y$-$\Gamma$-$M$} high symmetry path in the BZ marked in the inset, calculated for an applied magnetic field of \mbox{$B_{z}=30$ T}. The color scale bar encodes whether the bands have magnon-like, phonon-like, or mixed character. The bands are labelled 1--8 from the highest to the lowest energy. (b) The counterpart of panel (a), but in the absence of magnetoelastic coupling [$H_m=0$ in Eq.~\eqref{eq:mp}] and for zero applied  magnetic field [$B_z=0$ in Eq.~\eqref{eq:heisenberg}]. This means that red lines denote purely magnon bands and blue lines denote purely phonon bands of FePS$_3$, without any hybridization between them being present.}
  \label{Bands}
\end{figure}
The term $H_m$ is the anisotropic Heisenberg model \cite{Wildes2012,Lancon2016,Olsen2021,To2022,Lee2022}:
\begin{equation}\label{eq:heisenberg}
    H_{m} = \sum_{i,j} J_{ij}\boldsymbol{S}_{i}\boldsymbol{S}_{j} + \Delta\sum_{i}\left( S_{i}^{z}\right)^{2} + g\mu_{B}B_{z}\sum_{i}S_{i}^{z}
\end{equation}
where  $\boldsymbol{S}_{i}=(S_{i}^{x},S_{i}^{y},S_{i}^{z})$ is the operator of total spin localized at a site $i$ of the lattice; $J_{ij}$ is the exchange coupling  between localized spins at sites $i$ and $j$; $\Delta$ is the easy-axis anisotropy energy; the Zeeman (third on the right) term takes into account coupling to the applied magnetic field  $B_{z}$  pointing along the $z$-axis which is perpendicular to the plane in Fig.~\ref{Structure}; $g$ is the Land\'{e} $g$-factor; and $\mu_{B}$ is the Bohr magneton. The sum $\sum_{ij}$ runs over all atom pairs in the lattice up to the third-nearest neighbor. 

We take into account the magnetoelastic coupling by assuming that it acts only between magnons and out-of-plane phonons. Such assumption is particularly relevant for FePS$_3$ 2D AFM, where out-of-plane phonon modes are closely aligned with the magnon modes in terms of energy and have been observed to hybridize with them under an applied magnetic field~\cite{Liu2021}. Therefore, we focus only on the $z$-component of the lattice vibrations, so that describing them with a simple harmonic oscillator model yields the following effective phonon Hamiltonian~\cite{Zhang2019,Shen2020}
\begin{equation}\label{eq:hp}
    H_{p} = \sum_{i} \frac{\left(p_{i}^{z} \right)^{2}}{2M} + \frac{1}{2} \sum_{ij} u_{i}^{z} \Phi_{i,j}^{z}u_{j}^{z}.
\end{equation}
Here $p_{i}^{z}$ and $u_{i}^{z}$ are the operators of  out-of-plane momentum and displacement of the atom at site $i$ of the lattice, respectively; $\Phi^{z}$ is a spring constant matrix; and M is the mass of the atom. Finally, for the magnetoelastic coupling, which generates hybridization of magnon and phonon bands [Fig.~\ref{Bands}(a)], we adopt Hamiltonian derived by Kittel~\cite{Kittel1958} to linear order in the magnon amplitude, and adapted~\cite{Simensen2019,Liu2021} to magnons coupled to out-of-plane phonons in FePS$_3$
\begin{equation}\label{eq:mp}
    H_{mp} = -\xi\sum_{i}\left[ \epsilon^{yz}_{i} \left( S_{i}^{x}S_{i}^{z} + S_{i}^{z}S_{i}^{x} \right) + \epsilon^{xz}_{i} \left( S_{i}^{y}S_{i}^{z} + S_{i}^{z}S_{i}^{y} \right) \right],
\end{equation}
where $\xi$ is the coupling strength and $\epsilon^{xz}_{i}$ and $\epsilon^{yz}_{i}$ are strain functions at the $i$ site computed by averaging over the strain from nearest-neighboring ions
\begin{equation}
    \epsilon^{\alpha\beta}_{i} = \frac{1}{N}\sum_{j}\epsilon^{\alpha\beta}_{ij}.
\end{equation}
The two-ion strain tensor in the small displacement approximation is given by~\cite{Simensen2019,Sato2021}
\begin{equation}
    \epsilon^{\alpha\beta}_{ij}=\frac{1}{2} \left[ \left(r_{i}^{\alpha} - r_{j}^{\alpha} \right)\left( u_{i}^{\beta} - u_{j}^{\beta} \right) +\left(r_{i}^{\beta} - r_{j}^{\beta} \right)\left( u_{i}^{\alpha} - u_{j}^{\alpha} \right) \right],
\end{equation}
where $r_{i}^{\alpha}$ and $u_{i}^{\alpha}$ are the $\alpha$-component of the location vector in equilibrium and the displacement of the atom from equilibrium, respectively, for site $i$ of the lattice.


The transformation of Eq.~\eqref{Ham_total} into second-quantized notation is given in the Supplemental Material (SM)~\footnote[1]{See Supplemental Material at \url{https://mrsec.udel.edu/publications/}, which includes Refs.~\cite{Holstein1940,Lowdin1951,Lowdin1982,Jin2011,Schrieffer1966,Bravyi2011,Zhou2020,Wan2021,Massarelli2022,Ohashi2020}, for: ({\em i}) magnon-phonon Hamiltonian in the second-quantization formalism and details of its exact diagonalization; ({\em ii}) physical interpretation of the gap opening [Eq.~\eqref{eq:gap}] between two magnon-like bands energetically distant from anticrossing regions, as found in exact diagonalization, by using perturbation from phonons onto magnons examined via the {L\"{o}wdin}  partitioning; and ({\em iii}) additional details of Berry and spin Berry curvature calculations and related symmetry arguments}.  Since this Hamiltonian is quadratic in creation and annihilation operators for magnons and phonons, it can be exactly diagonalized to obtain quasiparticle band structure in Figs.~\ref{Bands} for magnon-polaron quasiparticle. For easy comparison, Fig.~\ref{Bands}(b) plots non-hybridized magnon (red curves) and phonon (blue curves) bands in the absence of magnetoelastic coupling [$H_m=0$ in Eq.~\eqref{eq:mp}] and for zero applied magnetic field [$B_z=0$ in Eq.~\eqref{eq:heisenberg}].

\subsection{Transverse thermal and spin transport in the linear response regime}\label{sec:conductivities}

Within the linear response theory, the equations describing transverse quasiparticle transport underlying THE and SNE are given by \cite{Matsumoto2011,Shindou2013a,Shindou2013b,Matsumoto2014,Li2020,Go2022}
\begin{align}
    j_{y}^{Q} =-\kappa_{xy}\partial_{x}T, \label{eq:thecon} \\
    j_{y}^{S^{z}} = -\eta_{xy}^{S^{z}}\partial_{x}T, \label{eq:snecon}
\end{align}
where $j_{y}^{Q}$ and $j_{y}^{S^{z}}$ are  thermal  and spin current, respectively,  flowing along the $y$-axis in  response to  the temperature gradient  $\partial_{x}T$ applied along the $x$-axis [Fig.~\ref{Structure}]. The coefficients of proportionality in Eqs.~\eqref{eq:thecon} and ~\eqref{eq:snecon} are the thermal Hall conductivity 
\begin{equation} \label{kappa}
    \kappa_{xy} = -\frac{k_{B}^{2}T}{\hbar} \sum_{n=1}^{N} \int F_{2} \left(\rho_{n} \right) \Omega_{n}^{z}d\boldsymbol{k},
\end{equation}
and the spin Nernst conductivity 
\begin{equation} \label{alpha}
    \eta_{xy}^{S^{z}} = \frac{k_{B}}{\hbar} \sum_{n=1}^{N} \int F_{1} \left(\rho_{n} \right) \Omega^{z}_{S^{z},n}d\boldsymbol{k}.
\end{equation}
Here $\rho_{n} = [e^{E_{n}/k_{B}T}-1]^{-1}$ is the Bose-Einstein distribution function, with $E_{n}$ being the eigenenergy of the $n$th band, which enters into the conductivity expressions through functions 
\begin{equation}\label{F1}
    F_{1}\left( \rho_{n}\right) = \left(1+ \rho_{n} \right)ln\left(1+ \rho_{n} \right) -\rho_{n}ln\left(\rho_{n} \right),
\end{equation}
or
\begin{equation}\label{F2}
    F_{2}\left( \rho_{n}\right) = \left(1+ \rho_{n} \right)\ln^{2}\left(1+ \frac{1}{\rho_{n}} \right) -\ln^{2}\left(\rho_{n} \right)-2\mathrm{Li}_{2}\left(-\rho_{n} \right),
\end{equation}
where $\mathrm{Li}_{2}$ is the polylogarithm function.  Finally, the Berry $\boldsymbol{\Omega}_{n}(\boldsymbol{k})$ and spin (generalized) spin Berry $\boldsymbol{\Omega}_{S^{\alpha},n}(\boldsymbol{k})$  curvature of the $n$th band are given by~\cite{Li2020,Go2022}
\begin{widetext}
\begin{align} \label{EqBerry}
   \boldsymbol{\Omega}_{n}\left(\boldsymbol{k} \right) = \sum_{m\neq n}\frac{i\hbar^{2}\langle n(\boldsymbol{k})\vert \boldsymbol{v} \vert m(\boldsymbol{k}) \rangle \langle m(\boldsymbol{k}) \vert {\bm \sigma}_{3} \vert m(\boldsymbol{k})\rangle \times \langle m(\boldsymbol{k})\vert \boldsymbol{v} \vert n(\boldsymbol{k}) \rangle \langle n(\boldsymbol{k}) \vert {\bm \sigma}_{3} \vert n(\boldsymbol{k})\rangle}{\left[ \sigma_{3}^{nn}E_{n} \left( \boldsymbol{k} \right) - \sigma_{3}^{mm}E_{m} \left( \boldsymbol{k} \right) \right]^{2}},
\end{align}
and
\begin{align}\label{EqSpinberry}
\boldsymbol{\Omega}_{S^{\alpha},n}\left(\boldsymbol{k} \right) = \sum_{m\neq n}\frac{i\hbar^{2}\langle n(\boldsymbol{k})\vert \boldsymbol{j}^{S^{\alpha}}\vert m(\boldsymbol{k}) \rangle \langle m(\boldsymbol{k}) \vert {\bm \sigma}_{3} \vert m(\boldsymbol{k})\rangle \times \langle m(\boldsymbol{k})\vert \boldsymbol{v} \vert n(\boldsymbol{k}) \rangle\langle n(\boldsymbol{k}) \vert {\bm \sigma}_{3} \vert n(\boldsymbol{k})\rangle}{\left[ \sigma_{3}^{nn}E_{n} \left( \boldsymbol{k} \right) - \sigma_{3}^{mm}E_{m} \left( \boldsymbol{k} \right) \right]^{2}},
\end{align}
\end{widetext}
where we use $E_{n}(\boldsymbol{k})$  and $\vert n (\boldsymbol{k})\rangle$ to denote the eigenvectors and eigenvalues, respectively, obtained from Colpa's diagonalization algorithm~\cite{Colpa1978,Colpa1979,Colpa1986a,Colpa1986b} (see the SM~\footnotemark[1] for details);  $\boldsymbol{v}=(v_x,v_y,v_z)$ denotes the velocity vector operator; $\boldsymbol{j}^{S^\alpha}$ denotes the spin current tensor operator 
\begin{equation}
    \boldsymbol{j}^{S^{\alpha}} = S^{\alpha}{\bm \sigma}_{3}\boldsymbol{v} + \boldsymbol{v}{\bm \sigma}_{3}S^{\alpha};
\end{equation}
and ${\bm \sigma}_{3}$ matrix is given by
\begin{equation}
    {\bm \sigma}_{3} = \begin{pmatrix}
    \boldsymbol{1}_{N \times N} & 0 \\
    0 & -\boldsymbol{1}_{N \times N}
    \end{pmatrix},
\end{equation}
where $\boldsymbol{1}_{N \times N}$ is  $N \times N$ identity matrix and \mbox{$\sigma_{3}^{nn}=\langle n(\boldsymbol{k})\vert {\bm \sigma}_{3}\vert n(\boldsymbol{k})\rangle$} is the $n$th diagonal element of ${\bm \sigma}_{3}$. Thus, evaluating Berry [Eqs.~\eqref{EqBerry}] and spin Berry [Eq.~\eqref{EqSpinberry}] curvatures directly yields the thermal and spin Nernst conductivities, respectively.

\section{Results and Discussion}\label{results}

\subsection{Topological transport of magnon-polarons: Thermal Hall and spin Nernst effects}

\begin{figure}
\centering
    \includegraphics[width=.46\textwidth]{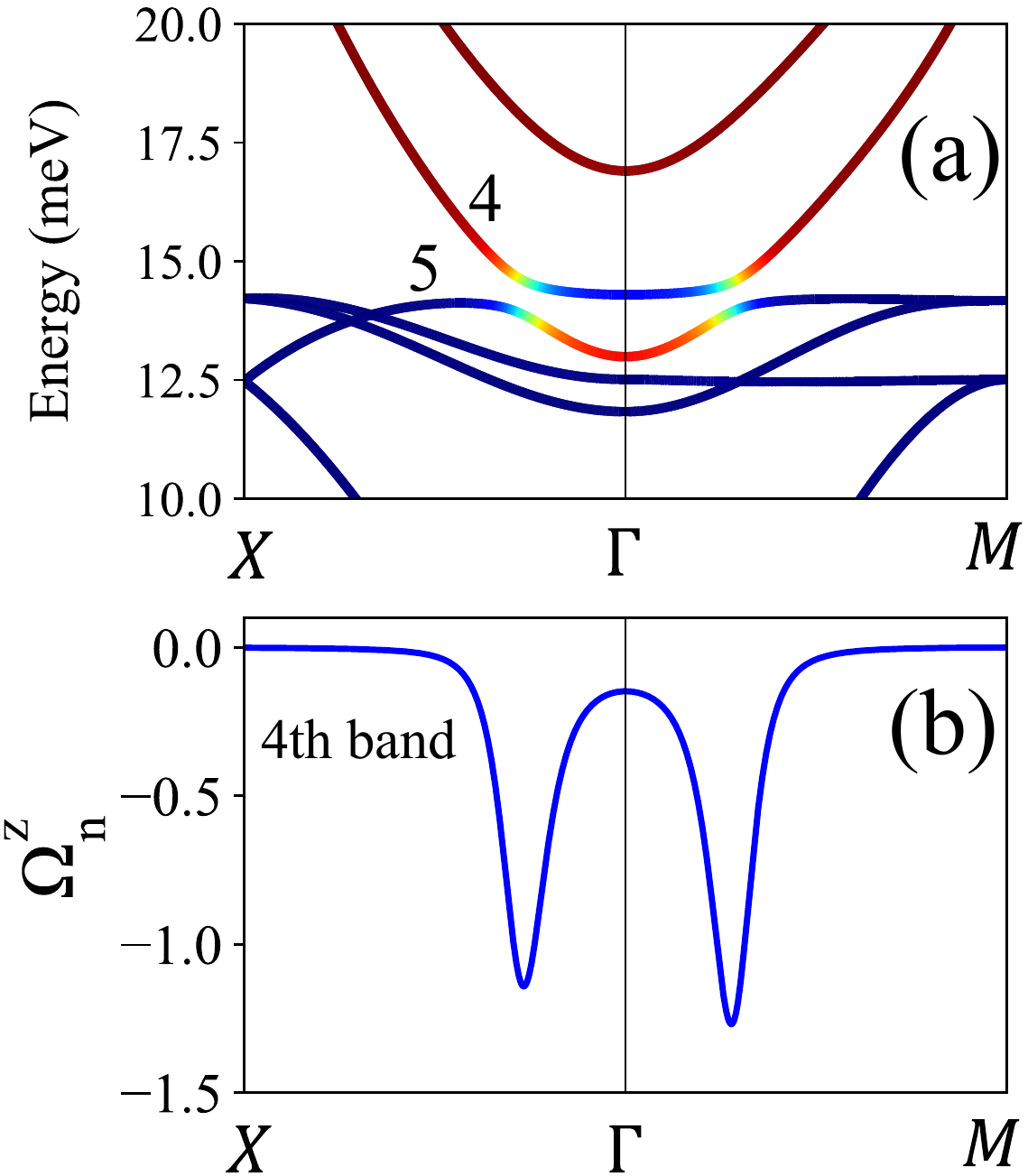}
 \caption{(a) The hybridized magnon-phonon band structure of 2D FePS$_3$, along $X$-$\Gamma$-$M$ high symmetry path, calculated for an applied magnetic field of \mbox{$B_{z}=30$ T}. (b) The corresponding Berry curvature $\Omega_{n}^{z}$ along the $X$-$\Gamma$-$M$ path calculated for the 4th band in panel (a).}
  \label{BandvsBerry}
\end{figure}

We first assume that FePS$_3$ is exposed to an applied  magnetic field of \mbox{$30$ T}. Figure~\ref{BandvsBerry}(a) show a zoom onto magnon-phonon hybridized bands from Fig.~\ref{Bands} focused on 4th (predominantly magnon, as it is mostly red) and 5th (predominantly phonon, as it is mostly blue) band in the energy window between 10 and 20 meV along the $X$-$\Gamma$-$M$ path. These two bands are strongly coupled, which results in  two  anticrossings [Fig.~\ref{BandvsBerry}(a)]. In the vicinity of these  anticrossings, the eigenstates are hybridized, $\psi_\mathrm{hybrid} = \psi_\mathrm{magnon} \pm \psi_\mathrm{phonon}$, with both magnon and phonon character. The presence of such  superpositions are denoted by  the bright green-yellow color of the bands in the anticrossing region [Fig.~\ref{BandvsBerry}(a)]. We note that {\em both} an applied magnetic field and magnetoelastic coupling between magnons and phonons are required for such hybridization and anticrossing to emerge---the magnetoelastic coupling provides the necessary interaction, while the magnetic field tunes the magnon and phonon bands toward energy degeneracy.

The hybridization of two distinct excitations leads to a finite Berry curvature. Let us recall that, e.g.,  hybridization of $s$- and $p$-states in HgTe/CdTe semiconductor quantum wells causes nontrivial topological properties for electrons at the Fermi level~\cite{Bernevig2006}. The physics here is analogous---in the region of the BZ where the magnon band (4th band) and phonon band (5th band) anticross we expect nonzero Berry curvature. In contrast, we expect that away from the anticrossing regions,  the Berry curvature should vanish because either band is dominated solely by magnon or phonon character. Figure~\ref{BandvsBerry}(b), showing the Berry curvature [Eq.~\ref{EqBerry}] for the 4th band along the same $X$-$\Gamma$-$M$ path, confirms this expectation as $\Omega^{z}_{n}\left(\boldsymbol{k} \right) \neq 0$ in Fig.~\ref{BandvsBerry}(b) only around the anticrossing regions  identified in Fig.~\ref{BandvsBerry}(a). Thus, the nontrivial topology of magnon and phonon bands in FePS$_3$ emerges due to their hybridization via magnetoelastic coupling [Eq.~\eqref{eq:mp}], while these bands individually [Fig.~\ref{Bands}] exhibit trivial topology [Fig.~\ref{BandvsBerry}(b)].

\begin{figure}[b!]
\centering
    \includegraphics[width=1.03\textwidth]{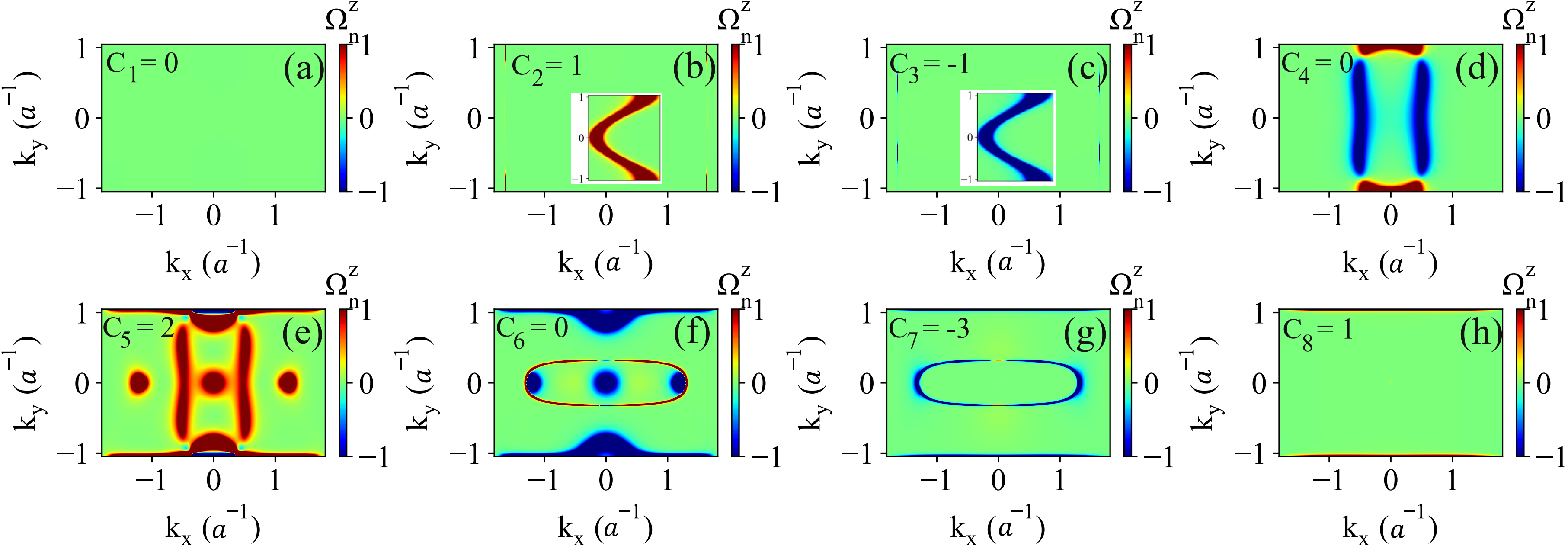}
 \caption{The Berry curvature $\Omega_n^z$ [Eq.~\eqref{EqBerry}] computed for magnon-phonon bands [Fig.~\ref{Bands}] of FePS$_3$ as a function of the in-plane wavevector $(k_{x},k_{y})$ within the first BZ and using  applied magnetic field \mbox{$B_{z}=30$ T}. Panels (a)--(h) correspond to bands 1--8 denoted in Fig.~\ref{Bands}. Their corresponding Chern number $C_{n}$ ($n=1,2,...,8$) in Eq.~\eqref{eq:chern} is provided in the upper left corner of each panel. The insets in panels (b) and (c) show a zoom in around $k_{x}=-1.64~a^{-1}$ where the Berry curvature of the corresponding bands is nonzero.}
  \label{Berry}
\end{figure}

Figure~\ref{Berry} shows the Berry curvature for the eight bands 1--8 in Fig.~\ref{Bands} as a function of the in-plane wave vector $\boldsymbol{k}=(k_{x},k_{y})$. In each panel, we also report the Chern number calculated as
\begin{equation}\label{eq:chern}
C_{n}=\frac{1}{2\pi}\int_{BZ}\Omega_{n}^{z} (\boldsymbol{k})dk_{x}dk_{y}.
\end{equation}
These calculations were performed for an applied magnetic field $B_{z}=30~T$ that causes the lowest magnon band to overlap with the out-of-plane optical phonon bands, as shown in Fig.~\ref{BandvsBerry}(a). Non-zero Berry curvature is observed in the vicinity of anticrossing regions in the 4th, 5th, and 6th bands in the color plot. The 1st band [Fig.~\ref{Berry}(a)] has zero Berry curvature everywhere, which obviously leads to zero Chern number. The 4th and 6th bands [Figs.~\ref{Berry}(d) and ~\ref{Berry}(f)] have non-zero Berry curvature, but the integral of the Berry curvature over the entire BZ of these bands vanishes. As a result, the Chern number is zero and these are topologically trivial bands. The other bands all have nonzero Chern number, with the sum of their Chern numbers obeying the sum rule, $\sum_{i=1}^{N} C_{i}=0$, as expected for a Bogoliubov-de Gennes (BdG) Hamiltonian~\cite{Park2020} (see the SM~\footnotemark[1] for more details on the BdG Hamiltonian construction). The bands with nonzero Chern number will contribute to THE via Eq.~\eqref{kappa}.

\onecolumngrid\
\twocolumngrid\

\vspace{0.3cm}

\onecolumngrid\
\begin{figure}[t!]
\centering
    \includegraphics[width=1.03\textwidth]{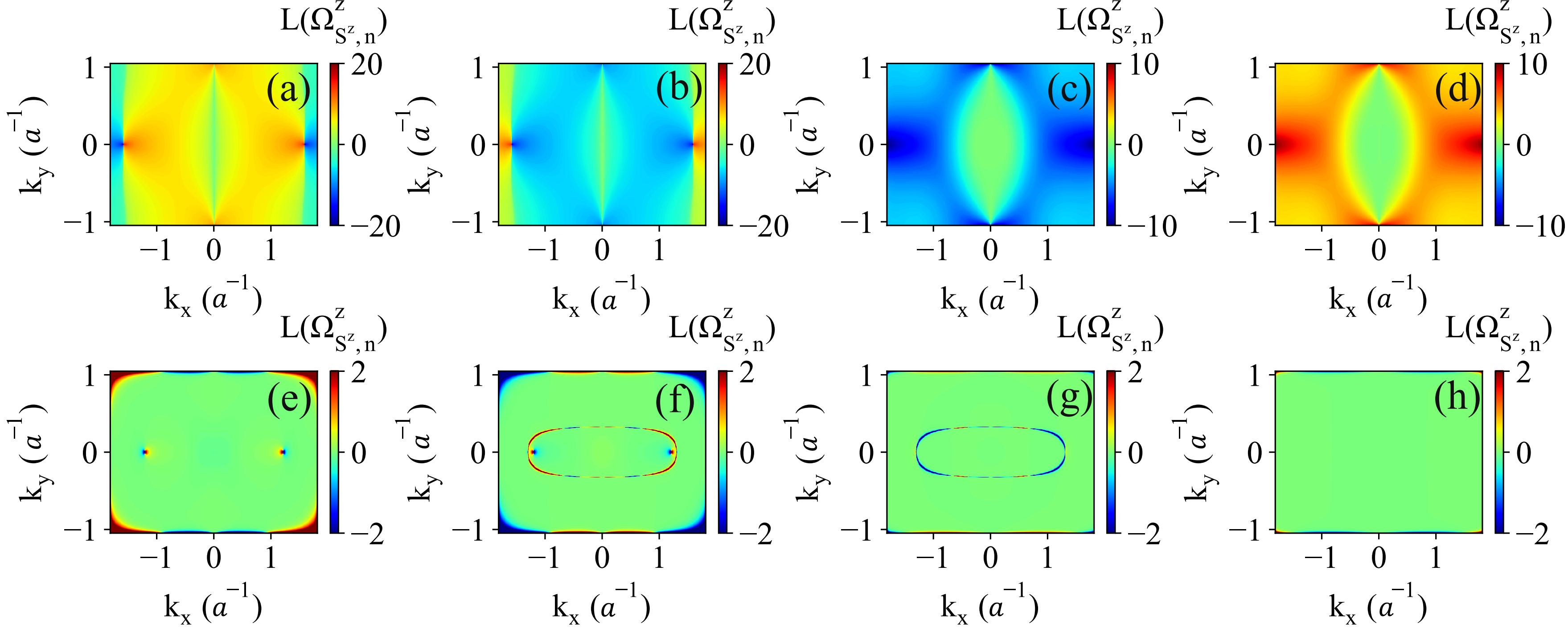}
 \caption{
The {\em spin} Berry curvature $\Omega_{S_{z},n}^{z}$  [Eq.~\eqref{EqSpinberry}] computed for magnon-phonon bands [Fig.~\ref{Bands}] of FePS$_3$ as a function of the in-plane wavevector $(k_{x},k_{y})$ within the first BZ and using in the absence of applied magnetic field $B_{z}=0$. Panels (a)--(h) correspond to bands 1--8 denoted in Fig.~\ref{Bands}.  The color bar encodes the magnitude of the function $L=\mathrm{sgn}(\Omega_{S^{z},n}^{z})\log(1+\vert \Omega_{S^{z},n}^{z}\vert)$.}
  \label{Spinberry}
\end{figure}
\twocolumngrid\

However, it is surprising and quite different from standard lore~\cite{Zhang2022,Park2020,Zhang2020,Bazazzadeh2021} that non-zero Berry curvature can be found for the 2nd [Fig.~\ref{Berry}(b)], 3rd [Fig.~\ref{Berry}(c)] and 8th [Fig.~\ref{Berry}(h)] band because these bands are well above or well  below the energy window in which magnon and phonon bands become degenerate in energy and anticross [Fig.~\ref{Bands}]. These bands all have non-trivial topology with a Chern number equal to $\pm 1$. The finite Berry curvature and nontrivial topological properties of these bands can be understood as follows. Magnetoelastic interaction facilitates coupling between magnon and phonon bands even when they are {\em not}  energetically close together, so that magnon bands have  small phononic character and vice versa~\cite{Go2022}. This effect can open a gap between two magnon-like bands [such as 2nd and 3rd in Figs.~\ref{Berry}(b) and \ref{Berry}(c)] at $k_{x} = \pm 1.64~(a^{-1})$, thereby making possible interband transitions between these two [see the inset of Fig.~S2(b) in the SM~\footnotemark[1] for details]. Without magnetoelastic coupling, these magnon bands are degenerate, i.e., they cross each other at $k_{x} = \pm 1.64~(a^{-1})$ [Fig.~S2(a) in the SM~\footnotemark[1]]. Precise quantum-mechanical interpretation of this picture can be obtained from the perturbation theory---the gap opening between the two magnon-like bands is due to perturbations from phonons, which appears as a second order correction term
    \begin{equation}\label{eq:gap}
    \delta E^{m}_{ij} \propto \sum_{p}\left[ \bar{H}\right]_{mi,p} \left[ \bar{H}\right]_{p,mj}\left[\frac{1} { \bar{E}_{mi} - \bar{E}_{p}} + \frac{1} { \bar{E}_{mj} - \bar{E}_{p}} \right]
\end{equation}
to the magnon band levels [for derivation of Eq.~\eqref{eq:gap} see the SM~\footnotemark[1]]. Here the indices $p$, $mi$, $mj$ indicate the phonon states which  mediate interband transitions between magnon states $i$ and $j$;  $\left[ \bar{H}\right]_{mi,p}$  ($\left[ \bar{H}\right]_{p,mj}$) describes the coupling between $i$ magnon (phonon) band and phonon ($j$ magnon) states; $\bar{E}_{mi}$, $\bar{E}_{mj}$ and $\bar{E}_{p}$ are eigenenergies of $i$ magnon, $j$ magnon, and phonon states, respectively, as obtained from exact diagonalization of the bosonic magnon-phonon Hamiltonian (see the SM~\footnotemark[1] for details). As the result, the Berry curvature of the 2nd and 3rd band at around $k_{x}=\pm 1.64~(a^{-1})$, which is associated with the tiny avoided crossing points between the 2nd and 3rd magnon-like bands, {\em becomes finite}. An analogous effect occurs for the phonon bands. For instance, a magnon-mediated phonon-phonon interband transition between 7th and 8th bands in Fig.~\ref{Bands}(a) generates a finite Berry curvature at $k_{y} \approx \pm 1~(a^{-1})$ for the 8th (phonon-like) band, as confirmed by Fig.~\ref{Berry}(h).

Another consequence of these phonon-mediated magnon-magnon and magnon-mediated phonon-phonon interband transitions is that they induce the topological transverse transport of spin angular momentum carried by magnons even at {\em zero} applied magnetic field. Figure~\ref{Spinberry} shows the computed spin (generalized) Berry curvature [Eq.~\eqref{Spinberry}] for bands 1--8 [Fig.~\ref{Bands}] calculated for $B_{z}=0$. We note that in the absence of both applied magnetic field and magnon-phonon coupling, the magnon bands exhibit a double degeneracy, with one set of bands carrying spin up [such as the 1st band in Fig.~\ref{Bands}(a)] and another set carrying spin down [such as the 2nd band in Fig.~\ref{Bands}(a)]. Consequently, the band structures of the magnon-phonon system in FePS$_3$ also exhibit a double degeneracy, as illustrated in Fig.~\ref{Bands}(b). However, the magnetoelastic coupling between the magnetic and elastic degrees of freedom in FePS$_3$ lifts the degeneracy of these two magnon bands with opposite spin, therefore making possible for interband transition between those two magnon-like bands of opposite spin, even in the absence an applied magnetic field (see the SM~\footnotemark[1] for Fig.~S3 and details of calculations). Such phonon-mediated interband transitions between magnon-like bands, which are energetically distant from usually considered~\cite{Zhang2022,Park2020,Zhang2020,Bazazzadeh2021} anticrossing regions [Fig.~\ref{BandvsBerry}(a)] of hybridized magnon-phonon bands, can result in a very large spin Berry curvature found in Fig.~\ref{Spinberry}(a)--(d) because of the smallness~\cite{Go2022} [with respect to the gap in anticrossing regions in Fig.~\ref{BandvsBerry}(a)] of energy gap  between the two magnon-like bands with opposite spin polarization [Fig.~S3(b) in SM~\footnotemark[1]]. The same effect can operate between phonon-like bands. For example, the 7th and 8th (phonon-like) bands in Fig.~\ref{Bands}(a) will exhibit magnon-mediated interband transitions, thereby developing finite spin Berry curvature [Figs.~\ref{Spinberry}(g) and ~\ref{Spinberry}(h)].   


\begin{figure}
\centering
    \includegraphics[width=.48\textwidth]{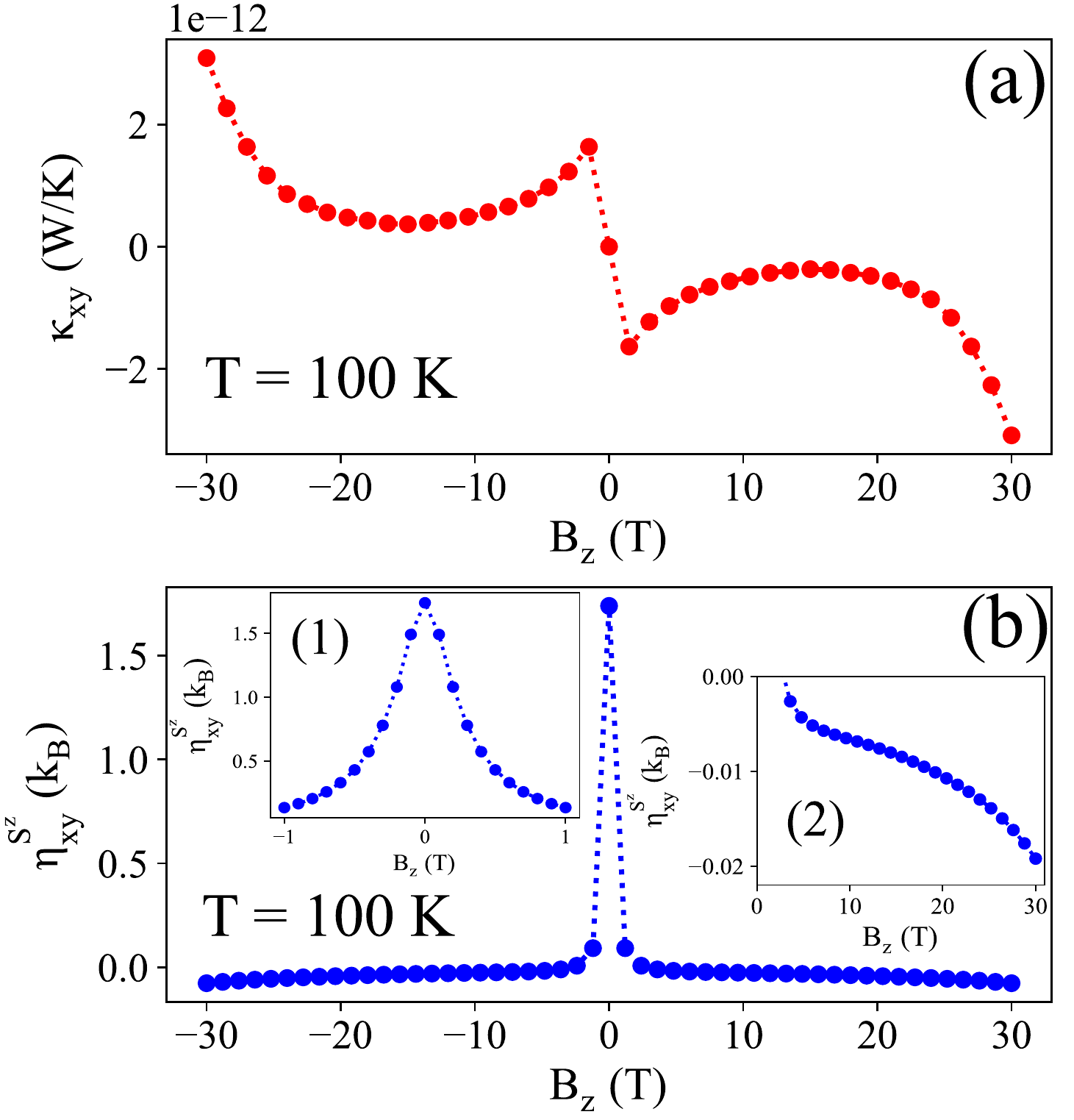}
 \caption{(a) Thermal Hall  and (b) spin Nernst conductivities as a function of an applied magnetic field $B_z$. These conductivities are calculated at \mbox{$T=100$ K} using  FePS$_3$ magnon-phonon band structure [Fig.~\ref{Bands}] and its Berry [Fig.~\ref{Berry}] and spin Berry [Fig.~\ref{Spinberry}] curvatures. Two insets in panel (b) show a zoom in 
 for: (1) $B_{z} \in [-1~\mathrm{T},1~\mathrm{T}]$; and (2) $B_{z} \in [2~\mathrm{T},30~\mathrm{T}]$.}
  \label{SNcondB}
\end{figure}


\subsection{Magnetic field dependence of the thermal Hall and spin Nernst effects on applied magnetic field}

Using computed Berry [Fig.~\ref{Berry}] and  spin Berry [Fig.~\ref{Spinberry}] curvatures, we can obtain directly thermal Hall [via Eq.~\eqref{kappa}] and spin Hall [via Eq.~\eqref{alpha}] conductivities shown in Figs.~\ref{SNcondB}(a) and  ~\ref{SNcondB}(b), respectively as a function of applied magnetic field at fixed temperature \mbox{$T=100$ K} that is below the N\'{e}el temperature of FePS$_3$. As expected, the thermal Hall conductivity changes sign when we reverse the  applied magnetic field, i.e., $\kappa_{xy}(B_{z})=-\kappa_{xy}(-B_{z})$. In the absence of applied magnetic field [$B_{z}=0$ point in Fig.~\ref{SNcondB}(a)], the thermal Hall conductivity vanishes. We can understand this feature by recognizing that when the applied magnetic field is absent, the system will be invariant under the time-reversal symmetry operation $\mathcal{T}$ combined with the spin rotation symmetry operation $\mathcal{C}$ that flips all spins in the system. The combination of these operations leads to an effective time reversal symmetry (TRS) operation $\mathcal{T'}=\mathcal{T C}$ under which $\partial_{x}T$ is preserved while the thermal Hall current is transformed as $j_{y}^{Q} \rightarrow -j_{y}^{Q}$. Because this system preserves $\mathcal{T'}=\mathcal{T C}$ symmetry, $j_{y}^{Q} = -j_{y}^{Q}=0$ and the thermal Hall conductivity $\kappa_{xy}$ must be zero. We note that even though the thermal Hall conductivity $\kappa_{xy}$ of the 
magnon-phonon hybridized system is zero in zero magnetic field, the Berry curvature  $\Omega_{n}^{z}(\boldsymbol{k})$ of individual bands may be finite at specific $k$-points within the BZ, as long as the integral of the Berry curvature over the entire BZ vanishes (see the SM~\footnotemark[1] for a detailed argument). This ensures that THE induced by the magnon-phonon hybridization does not occur without breaking the effective TRS~\cite{Zhang2020}.

\begin{figure}
\centering
    \includegraphics[width=.44\textwidth]{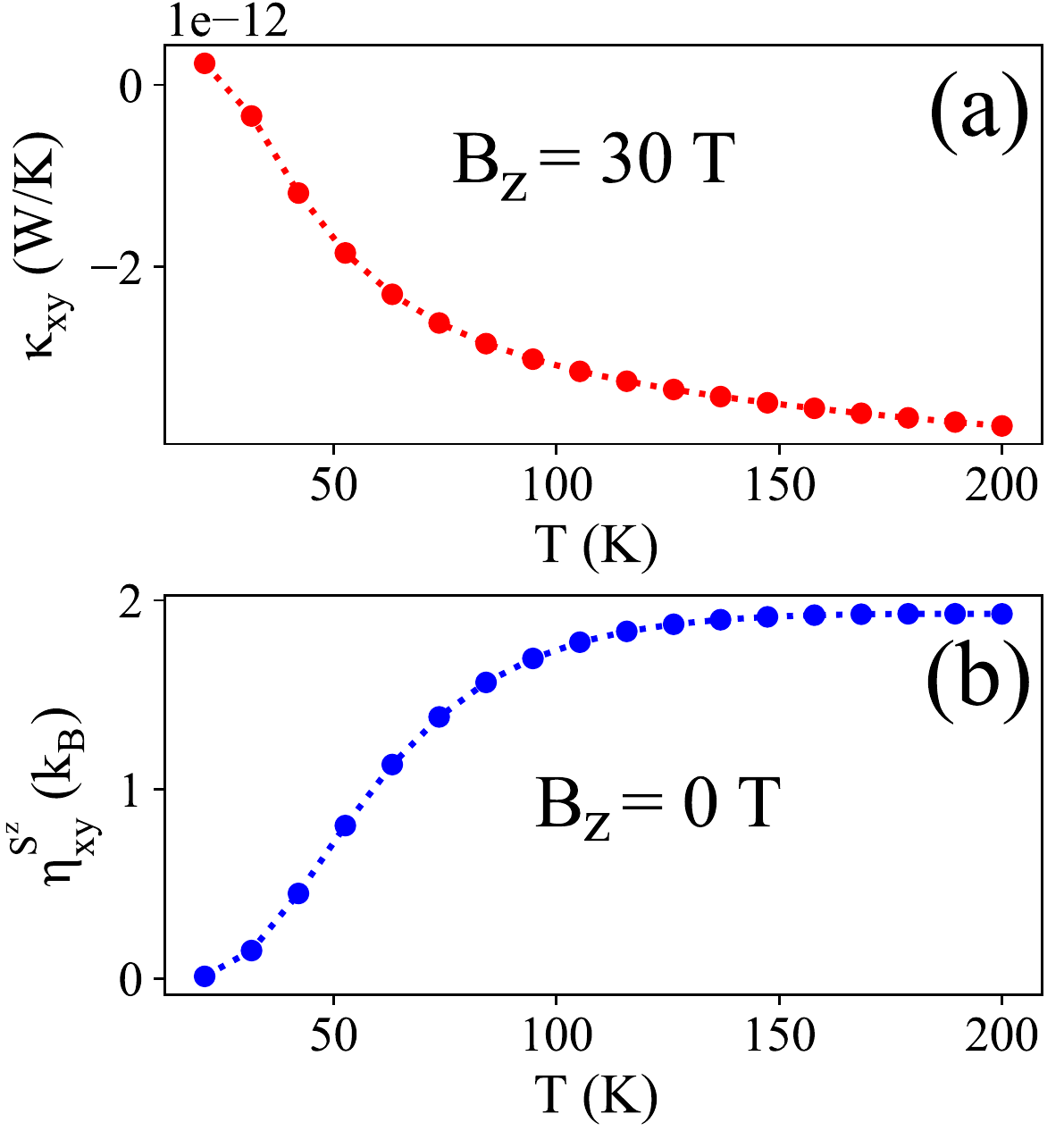}
 \caption{(a) Thermal Hall and (b) spin Nernst conductivities of FePS$_3$ as a function of temperature $T$ calculated for applied magnetic field \mbox{$B_{z}=30$ T} in (a) or \mbox{$B_{z}=0$ T} in (b).}
  \label{SCvsT}
\end{figure}

In contrast to the thermal Hall conductivity, the spin Nernst conductivity shown in Fig.~\ref{SNcondB}(b) is an even function of $B_{z}$, i.e., $\eta_{xy}^{S^{z}}(B_{z})=\eta_{xy}^{S^{z}}(-B_{z})$. Moreover, spin Nernst conductivity  be finite even in the absence of an applied magnetic field~\cite{Go2022}, i.e., under the effective time reversal symmetry $\mathcal{T'}$. Indeed, if we rewrite the thermal spin current [Eq.~\eqref{eq:snecon}]  as  $j_{y}^{S^{z}}=j_{y}^{S^{z \uparrow}}-j_{y}^{S^{z \downarrow}}$, then under  $\mathcal{T'}$ operation the spin-polarized currents on the right side  change the sign and flip the spin, i.e.,  $\mathcal{T'}j_{y}^{S^{z \uparrow}}=-j_{y}^{S^{z \downarrow}}$ and  $\mathcal{T'}j_{y}^{S^{z \downarrow}}=-j_{y}^{S^{z \uparrow}}$. This leads to $\mathcal{T'}j_{y}^{S^{z}}=-j_{y}^{S^{z \downarrow}} + j_{y}^{S^{z \uparrow}}=j_{y}^{S_{z}}$, which is always true because our system preserves the effective time reversal symmetry in the absence of an  applied magnetic field. It is therefore possible for the spin Nernst conductivity to be nonzero at zero applied magnetic field, as confirmed in  Fig.~\ref{SNcondB}(b). At zero or small applied magnetic field, the {\em giant} spin Nernst conductivity is mainly governed by phonon-mediated interband transitions between magnon-like bands. It then decays rapidly [inset (1) in Fig.~\ref{SNcondB}] when the applied magnetic field is \mbox{$B_z \gtrsim  2$ T}, dropping eventually by two orders of magnitude, because 
the energy spacing between the two magnon-like  bands increases and thus interband transitions between the two are suppressed. 

As the applied magnetic field magnitude increases from $\approx$ 2 to 30 T the spin Nernst conductivity slightly changes while becoming negative, $\eta_{xy}^{S^z}<0$ [inset (2) in Fig.~\ref{SNcondB}]. We find that from $\approx$ 2 to $\approx 5~$T, the spin Nernst conductivity originates primarily from magnon-mediated interband transitions between phonon-like bands. Once the phonon bands start hybridizing with magnon bands at \mbox{$B_{z} \approx 5$ T}, spin Berry curvature [Fig.~\ref{BandvsBerry}] at anticrossing regions of magnon-phonon bands also contribute, as amply explored in prior literature~\cite{Zhang2022,Park2020,Zhang2020,Bazazzadeh2021}. To understand why the spin Nernst conductivity becomes more negative with increasing applied magnetic field, we consider that in the conserved spin approximation the spin Nernst conductivity derived 
from semi-classical theory is given by \cite{Matsumoto2011,Cheng2016,Park2020}:
\begin{equation}\label{eq:semiclassicaeta}
    \eta_{xy}^{S^{z}} = -\frac{k_{B}}{\hbar V} \sum_{k}\sum_{n=1}^{N} \langle S^{z} \rangle \Omega_{n}^{z} F_{1}(E_{n}/k_{B}T)
\end{equation}
where $\langle S^{z} \rangle$ is the expectation value of $S^{z}$ operator, $\Omega_{n}^{z}$ is the Berry curvature of the $nth$ band and the $F_{1}$ function was defined in Eq.~\ref{F1}. From Eq.~\eqref{eq:semiclassicaeta}, we see that increasing applied magnetic field leads to both larger spin polarization and stronger hybridizations between magnon and phonon states due to the shift toward energy degeneracy of the magnon and phonon states. Consequently, the amplitude of the spin Nernst conductivity $\eta_{xy}^{S^{z}}$ is augmented within this regime.

Since the computed spin Nernst conductivity of FePS$_3$  around zero applied magnetic field is two orders of magnitude [Fig.~\ref{SNcondB}] larger than at $B_{z} \approx 10~$T, it should be possible to experimentally probe this effect by sweeping magnetic field. Furthermore, we note that spin Nernst conductivity of FePS$_3$ is much larger than that of other recently investigated 2D transition phosphorus trichalcogenides materials, such as MnPS$_3$, NiPS$_3$, NiPSe$_3$. Specifically, for FePS$_3$ in the zigzag phase studied here, the  computed spin Nernst conductivity is about four orders of magnitude larger than that of MnPS$_3$ in the N\'{e}el phase~\cite{Bazazzadeh2021}.  

We also emphasize that in the absence of magnetoelastic coupling, both the thermal Hall and spin Nernst conductivities vanish, irrespective of the applied magnetic field. This is because the system without magnetoelastic coupling preserves  $\mathcal{T}_{a}\mathcal{M}_{y}$ symmetry, where $\mathcal{M}_{y}$ is the mirror symmetry about the plane normal to the $y$-axis and $\mathcal{T}_{a}$ is a translation operator that moves the system by the vector $\boldsymbol{\beta}_{2}$ [Fig.~\ref{Structure}]. Unlike the effective time reversal symmetry $\mathcal{T'}$, $\mathcal{T}_{a}\mathcal{M}_{y}$ does not change the spin direction but does change the sign of both the thermal Hall and thermal spin Nernst current. In other words, one must have $j_{y}^{Q} = -j_{y}^{Q}=0$ and $j_{y}^{S^{z}}=-j_{y}^{S^{z}}=0$, therefore both the thermal Hall and spin Nernst conductivity must be zero. It is only when the magnetoelastic interaction breaks $\mathcal{T}_{a}\mathcal{M}_{y}$ symmetry that one obtains finite topological transverse transport of quasiparticles and their spin in a 2D AFM material.

Finally, Fig.~\ref{SCvsT} shows the thermal Hall and spin Nernst conductivities as a function of 
temperature using \mbox{$B_{z}=30$~T} or \mbox{$B_{z}=0$} applied magnetic field, respectively.  Both conductivities increase in magnitude with increasing temperature because there are increasing contributions to Berry and spin Berry curvature from phonon and magnon bands at higher energy. They start to  saturate at \mbox{$T \simeq 100$ K} when all  magnon bands at higher energy have already been included. We note that when \mbox{$T \simeq 0$ K}, the spin Nernst conductivity is almost zero, while the thermal Hall conductivity changes from positive to negative. This is because at very low temperature the main contributions to the THE come from the acoustic phonon band [8th band in Fig.~\ref{Bands}(a)] with positive Chern number $C_{8} = 1$ [Fig.~\ref{Berry}(h)]. As the temperature increases even slightly, the other bands with negative Chern number begin to contribute to topological transverse transport of quasiparticle and, thus, the thermal Hall conductivity becomes negative. In contrast, even though the spin Berry curvature of the  lowest phonon-like band [8th band in Fig.~\ref{Bands}(a)] is finite, the sum of the spin Berry curvature of the 8th band over the entire BZ vanishes to yield $\eta_{xy}^{S^z} \rightarrow 0$ at zero temperature. 

\section{Conclusions}\label{conclu}

In conclusion, we have investigated the transverse topological transport of magnon-polaron quasiparticles in the zigzag phase of FePS$_3$ 2D AFM. While we reproduce previous findings~\cite{Zhang2022,Park2020,Zhang2020,Bazazzadeh2021}, obtained for different realizations of 2D AFMs, on  magnetoelastic coupling mechanism where 
anticrossing regions of hybridized magnons-phonon bands provide key contribution~\cite{Park2020} to THE and SNE, we also predict  giant spin Nernst current carried by magnons even in zero applied magnetic field. This surprising finding was noticed before~\cite{Go2022}, but here we explain it thoroughly by using perturbative Eq.~\eqref{eq:gap} which reveals principal contribution to the spin Berry curvature behind SNE coming from interband transition between slightly gapped magnon-like bands that are far away in energy from usually considered anticrossing regions~\cite{Zhang2022,Park2020,Zhang2020,Bazazzadeh2021}. 
Of relevance to experimental probing of THE and SNE, which are currently lacking~\cite{Zhang2022}, our analysis indicates that FePS$_3$ will exhibit sizable thermal Hall conductivity and giant spin Nernst conductivities at temperatures of \mbox{$T \simeq 100$  K}, which is still below its N\'{e}el temperature  \mbox{$T_N \approx 118$ K}~\cite{Basnet2021,Liu2021}.

\begin{acknowledgments}
This research was primarily supported by NSF through the University of Delaware Materials Research Science and Engineering Center, DMR-2011824.
\end{acknowledgments}

\bibliography{biblio}


\end{document}